\documentstyle[11pt,newpasp,twoside,epsf]{article}
\newcommand{\um}{$\,\mu$m}
\newcommand{\bra}{Br~$\alpha$}

\def\edcomment#1{\iffalse\marginpar{\raggedright\sl#1\/}\else\relax\fi}
\reversemarginpar

\begin{document}
\title{Results from the South Pole InfraRed EXplorer Telescope}
 \author{J. M. Rathborne}
\affil{Institute for Astrophysical Research, 725 Commonwealth Avenue, Boston 
University, Boston, MA 02215, USA\\ School of Physics, University of
New South Wales, Sydney, NSW 2052, Australia}
\author{M. G. Burton}
\affil{School of Physics, University of New South Wales, Sydney, NSW 2052, Australia}

\begin{abstract}
The SPIREX telescope, located at the Amundsen--Scott South Pole
Station, was a prototype system developed to exploit the excellent
conditions for IR observing at the South Pole. Observations over two
winter seasons achieved remarkably deep, high-resolution, wide-field
images in the 3--5\um\ wavelength regime. Several star forming
complexes were observed, including NGC 6334, Chamaeleon I, $\eta$
Chamaeleontis, the Carina Nebula, 30 Doradus, RCW 57, RCW 38, as well
as the Galactic Centre. Images were obtained of lines at 2.42\um\
H$_{2}$, 3.29\um\ PAH and 4.05\um\ \bra, as well as 3.5\um\ L--band
and 4.7\um\ M--band continuum emission. These data, combined with
near--IR, mid--IR, and radio continuum maps, reveal the environments
of these star forming sites, as well as any protostars lying within
them. The SPIREX project, its observing and reduction methods, and
some sample data are summarized here.
\end{abstract}

\section{Introduction}

The South Pole InfraRed EXplorer (SPIREX) telescope was a prototype
system, developed to test the feasibility of building, operating, and
maintaining an infrared (IR) telescope during an Antarctic winter. The
initial driver for the SPIREX project was to exploit the conditions at
the South Pole that make it an excellent site for 3--5\,\um\,
observations --- that is the high altitude, the low temperatures, low
precipitable water vapour content in the atmosphere, and the stable
weather conditions (Burton et al.\ 1994; Marks et al.\ 1996; Hidas et
al.\ 2000; Marks 2002).

The SPIREX 60\,cm telescope began operations at the Amundsen--Scott
South Pole Station in 1994 (Hereld et al.\ 1990). SPIREX was initially
used as part of a campaign to measure the South Pole's thermal
background, sky transparency, and the fraction of time useful for IR
observations (see e.g.\ Ashley et al.\ 1996, Nguyen et al.\ 1996,
Phillips et al.\ 1996 and Chamberlain et al.\ 2000 for further
details on the IR sky conditions).

Due to the very low thermal background, SPIREX could perform longer
integrations in the 3--5\um\ regime, compared to temperate latitude
facilities. In addition to continuum emission which, at these
wavelengths pinpoints young embedded objects, there are many
astrophysically significant molecular lines accessible; e.g.\ those
from hydrogen and PAH molecules. Observations at 3--5\,\um\, are
difficult at most temperate sites, making studies involving these
lines limited, and thus the SPIREX dataset unique.

From 1994--1997 SPIREX was equipped with the GRIM (grism imager)
camera. This detector contained 128$\times$128 pixels and was
sensitive from 1--2.5\um. In 1998 the telescope was equipped and with the
Abu camera, which incorporated an engineering grade 1024$\times$1024
Aladdin detector (Fowler et al.\ 1998). This camera was sensitive from
2.4--5\um, and provided a 10\arcmin\, field of view image with a
0.6\arcsec\, pixel scale. Six science filters were available with
SPIREX/Abu. The three narrow-band filters were optimized to isolate
emission from molecular hydrogen (H$_{2}$ at 2.42\um), polycyclic
aromatic hydrocarbons (PAHs at 3.29\um), and hydrogen line emission
(\bra\, at 4.05\um). The three broad-band filters covered the L--,
L'-- and M--bands. The filter parameters and achieved sensitivities
are listed in Table~1 (see Burton et al., 2000).

\begin{table*}[t]
\caption{Parameters of the filters and achieved sensitivities for SPIREX/Abu.}
\vspace{0.1cm}
\begin{centering}
\begin{tabular}{|l|ccc|c|c|}
\hline
Filter       & Centre & Width  &   Range        & Time $\times$ Coadds\,$^{b}$ & Sensitivity$^{c}$ \\
             & \um  & \um  &   \um              &   & 3$\sigma$, $5'' \times 5''$, 1 hour \\
\hline
H$_{2}$$^{a}$& 2.425  & 0.034  &  2.408--2.441  & 360 $\times$ 1 & $\rm 6 \times 10^{-18}\, W m^{-2}$ \\
PAH          & 3.299  & 0.074  &  3.262--3.336  & 60 $\times$ 3 & 1 mJy\\
\bra         & 4.051  & 0.054  &  4.024--4.078  & 10 $\times$ 18 & $\rm 5 \times 10^{-17} \,W m^{-2}$\\
\hline
L            & 3.514  & 0.618  &  3.205--3.823  & 6 $\times$ 30 & 14.6 mags \\
L'           & 3.821  & 0.602  &  3.520--4.122  & 6 $\times$ 30 & 14.6 mags \\
narrow M     & 4.668  & 0.162  &  4.586--4.749  & 1.2 $\times$ 90 & 10.2 mags \\
\hline
\end{tabular}\\
\end{centering}
$^{a}$\,{\footnotesize{Covering the (1--0) Q(1)--Q(5) lines.}}\\
$^{b}$\,{\footnotesize{Typical integration time (in seconds) $\times$ number 
of coadded frames per position.}} \\
$^{c}$\,{\footnotesize{Achieved sensitivities were up to 1 magnitude worse than theoretical
sensitivities for the site with optimal instrument performance.}}\\
\end{table*}

SPIREX/Abu was well suited for studies of star forming
complexes. Young stars containing circumstellar disks have a colour
excess in the IR due to the absorption and re-emission of radiation
from the central star by the surrounding material. Recent studies have
found that the L--band may be the optimal wavelength for detection of
star and disk systems. Compared to the (H--K) colour
[$\equiv$~1.6--2.2\um], the (K--L) colour [$\equiv$~2.2--3.5\um] is
more sensitive to the presence of a disk (Haisch et al.\ 2000; Lada et
al.\ 2000; Kenyon \& Hartmann 1995).

Observations of PAH molecular line emission across star forming
complexes, allows one to study their environments. The fluorescent
emission from PAH molecules trace regions (known as photodissociation
regions or PDRs), where stellar UV radiation is heating the molecular gas
(Hollenbach \& Tielens 1997). PAH emission delineates externally
heated molecular clouds and reveals the interactions between nearby
massive stars and any remnant molecular material.

The first astronomical results from SPIREX were obtained when
Shoemaker-Levy 9 collided with Jupiter in 1994. Using the GRIM camera
at a wavelength of 2.36\um, images were obtained at 5 minute intervals
and captured 16 of the fragments, showing evidence of impact with
Jupiter in 10 cases (Severson 2000).  An extended halo around the
edge-on spiral galaxy ESO\,240--G11 was also imaged using GRIM at
2.4\um\ (Rauscher et al., 1998), reaching a sensitivity level of 25
mags/arcsec$^2$.

The following sections discuss a sample of data obtained from the two
years of operation of SPIREX/Abu (1998--99) at the South
Pole. Observations were conducted toward a number of different
complexes. The characteristics of these range from young to old, low-
to high-mass, and near and far star forming complexes. In particular,
results are presented here for NGC~6334, Chamaeleon~I,
$\eta$~Chamaeleontis, the Carina Nebula, 30~Doradus, RCW~57, RCW~38,
and the Galactic Centre.  In addition, SPIREX was used to search for
an infrared counterpart to the gamma-ray burst GRB990705 at 3.5\um\
(Masetti et al., 2000), though no source was detected to a limit of
13.9 magnitudes after 2 hours of integration.

\section{Data Acquisition and Reduction}

For each source position, a series of frames were obtained at the
specified integration time and then averaged. These parameters varied
depending on the observing wavelength, the properties of the filter
(narrow- or broad-band) and the weather conditions. Typical values for
each filter are given in Table~1. All observations were conducted by
the winter-over scientists at the South Pole station; Rodney Marks
(1998) and Charlie Kaminsky (1999).

The sequence for all observations consisted of a set of sky frames
followed by two sets of object frames. Each set consisted of five
averaged frames offset by {\mbox {$\sim$ 30\arcsec}} from the
previous. This sequence was repeated allowing the easy removal of sky
emission and artifacts from the array. Archived images were used for
dark subtraction and flat-fielding. Observations of standard stars
were obtained before and after all on-source observations for flux
calibration.

The majority of the data presented here was reduced by Joel Kastner
using the SPIREX/Abu data pipeline\footnote{See
http://pipe.cis.rit.edu/} (the exceptions are NGC 6334 and the
Carina Nebula). The data pipeline was a joint project of the Rochester
Institute for Technology Center for Imaging Science (RIT CIS), the
National Optical Astronomy Observatories (NOAO) and the Center for
Astrophysical Research in Antarctica (CARA).

To cover the star forming complexes of NGC 6334 and the Carina Nebula,
many adjacent positions were observed. Common stars in adjacent frames
were used to align the images and create the final larger mosaic
(using IRAF routines written by Peter McGregor\footnote{See
http://www.mso.anu.edu/observing/2.3m/CASPIR/}). The final PAH--band
mosaiced image for the Carina Nebula contained a total of 72
individual images, and for NGC 6334 the mosaic PAH--band image
contained 304 images.

\section{Results and Discussion}
\label{results-section}

\subsection{NGC 6334}

NGC 6334 is a young massive star forming region at a distance of
1.7\,kpc. It contains seven distinct sites of ongoing star formation
along a central molecular ridge. Fig.~1(a) shows the 3.29-\um\, data 
toward NGC 6334. 

The emission within this band comprises both 
PAH and continuum emission. A massive embedded star is seen at each
of several sites of star formation along the ridge, each $\sim 1$\,pc
apart.  They are surrounded by complex loops and filaments of PAH
emission. These PDRs have formed on the edge of the remnant molecular
cloud, as radiation from the young stars carves the material and ionizes
their surroundings. Analysis of the PAH emission and PDR features
across the central ridge of NGC 6334 have been presented in Jackson et
al.\ (1999) and Burton et al.\ (2000). The data shown here expand on
these studies, and further reveal the PDR structure adjacent to the
central star forming ridge.

In addition to the PAH--band images shown here, L--band images were
also obtained across NGC 6334.  When combined with near--IR data from
the 2MASS point source catalog they allow us to produce near--IR
colour--colour diagrams (e.g., (J--H) vs.\,(H--K) and (J--H)
vs.\,(K--L) diagrams). Results presented in Rathborne et al.\ (2003)
find 11 sources with a large IR excess using the (K--L) colour,
compared to just a single source using the (H--K) colour excess,
confirming that the (K--L) colour is far more sensitive to the
detection of circumstellar disks.

\begin{figure}[p]
\plotone{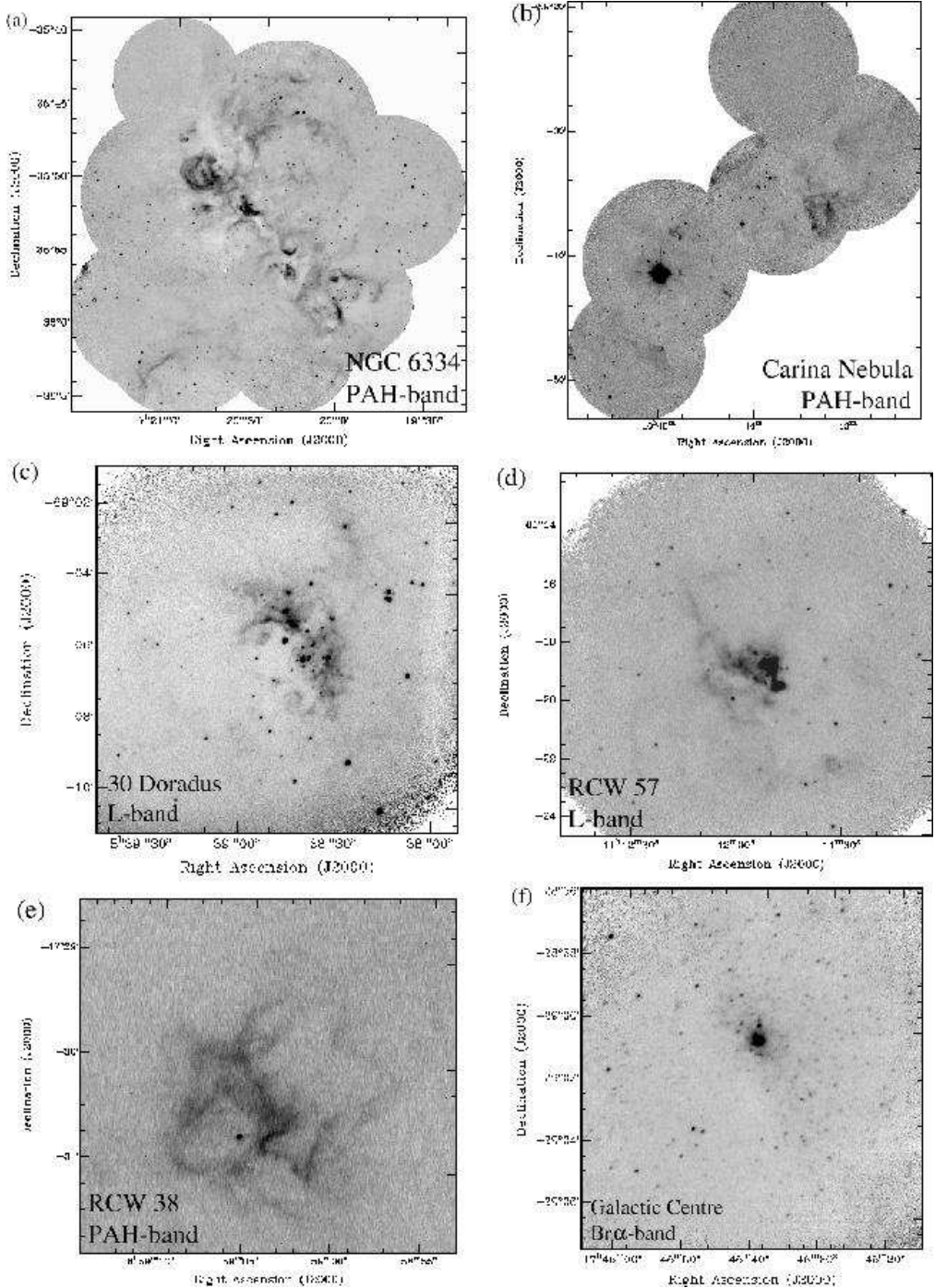}
\caption{A collection of the images obtained with SPIREX/Abu. 
(a) NGC 6334 in 3.29\um\ PAH emission (Burton et al., 2000, Rathborne
et al., 2003), (b) The Carina Nebula in PAH emission (Rathborne et
al., 2002), (c) 30 Doradus in 3.5\,\um\ L--band emission, (d) RCW 57
in L--band, (e) RCW 38 in PAH emission and (f) the Galactic Centre in
the 4.05\um\ Br$\alpha$ filter.  The RCW 38 and 57 images were kindly
provided by Chris Wright.}
\end{figure}

\subsection{Chamaeleon I}

Containing in excess of 100 pre-main sequence stars, the Chamaeleon I
dark cloud is one of the most active regions of nearby low-mass star
formation.  The central 0.5\,deg\,$^{2}$ of the complex was observed
in the L--band using SPIREX/Abu. These images reveal all of the known
pre-main sequence stars (to an L $\le$ 11).

Kenyon \& \& G{\' o}mez (2001) combined JHK observations obtained at
the CTIO with the SPIREX/Abu data, to construct near--IR
colour--colour diagrams. They find the fraction of sources with an IR
excess to be 58 $\pm$ 4\% (complete to an L $<$ 11). In addition, they
also confirm that sources with an IR excess are more easily identified
when using the (K--L) colour than (H--K).

\subsection{$\eta$ Chamaeleontis}

The $\eta$~Chamaeleontis cluster is one of the nearest to the Sun,
lying just 97\,pc away. It is also of intermediate age for a pre-main
sequence system, $\sim 9$\,Myrs old, and somewhat older than a number
of other clusters where the formation and evolution of circumstellar
disks has been studied.  Importantly, its proximity and compactness
mean that there is a complete population census for its members --- 15
stars ranging from 0.2 to 3.4\,M$_{\odot}$ in mass.  Lyo et al.\
(2003) imaged the cluster using SPIREX/Abu in the L--band, finding
60\% of the stars had IR excesses attributable to the presence of
disks around them.  Of those with disks, half showed a clear relation
between the strength of the IR excess at 3.5\um\ and the equivalent
width of the H$\alpha$ line emission, implying continuous accretion.
The lifetime of the disks in $\eta$~Chamaeleontis ($\sim$ 9\,Myrs) is
significantly longer than in a number of other systems that have been
studied, with lifetimes of 3--6\,Myrs (e.g., Haisch et al., 2001).

\subsection{The Carina Nebula}

The Carina Nebula is a massive star forming region at a distance of
2.2\,kpc. Because no protostellar objects had been found within this
nebula, it was thought to be `evolved' and devoid of current star
formation activity (e.g., Cox \& Bronfman 1995).  It does however,
contain 53 O--type stars and includes the clusters Tr 14 and Tr 16. At
its centre lies the massive star $\eta$ Car and the Keyhole
Nebula. Optical and near--IR images show a high amount of extinction
toward this complex, with many dark lanes, globules, clumps, and
filaments. Two large molecular clouds are located at the edges of the
complex, in close proximity to the massive stars.

Fig.~1(b) shows the PAH--band emission across the central Carina
Nebula (these data are also discussed in Brooks et al.\ 2000 and
Rathborne et al.\ 2002). When combined with 2.12--\um\ H$_{2}$ line
emission, MSX 8-- and 21--\um\ images, SEST molecular line maps, and
MOST 843\,MHz radio continuum images, three different environments are
revealed across the complex: (i) the Keyhole nebula containing
discreet, dense molecular clumps with PDRs on their surfaces; (ii)
PDRs at the edges of both the southern and northern molecular clouds;
and (iii) heated dust, intermixed with the PDRs surrounding Tr 14 and
the northern molecular cloud. Several 3.29--, 8-- and 21--\um\ point
sources were also located across the complex, with spectral energy
distributions corresponding to compact H\,II regions. Interestingly,
these were all found on the edges of PDRs which appear to have been
carved out by the stellar winds and radiation from the nearby massive
stars. These results suggest that star formation is indeed ongoing
within the Carina Nebula and may in fact be triggered by the
interactions resulting from the nearby massive stars.

\subsection{30 Doradus}

The 30 Doradus region is the brightest H\,II region in the Large
Magellanic Cloud. It contains a central cluster of massive stars
surrounded by extended nebulosity in the near--IR, with many
protostellar candidates. To study the nature of the embedded stellar
population in more detail, L--band observations were obtained with
SPIREX/Abu. The data presented in Fig.~1(c) are the most sensitive
ever obtained of the 30 Doradus complex.  The faintest start seen in
the image, which required 9.25 hours of on-source integration, has L =
14.5 magnitudes.  Many sources with an IR excess are revealed when the
L--band data are combined with JHK observations.  Several embedded
massive stars are apparent and in addition, a population census of the
young stellar objects can be conducted.  This still remains the deepest,
wide-field L--band image ever obtained, despite the small size of the
telescope, with an extended source sensitivity ($1 \sigma$) of 18.2
magnitudes per square arcsecond.

\subsection{RCW 57}

RCW 57 is a bright southern H\,II region. It contains a tight cluster
of massive stars and shows extended infrared nebulosity. L--band
observations of the central 10\arcmin\, of this complex were obtained
with SPIREX/Abu (Fig.~1(d)). This image reveals many point sources, in
addition to both bright and diffuse extended emission. Using JHK and
the SPIREX/Abu L--band data, studies are currently underway into the
star formation history of this complex.

\subsection{RCW 38}

RCW 38 is also a bright H\,II region containing a tightly packed
cluster of stars. Near--IR images of this complex show extended
nebulosity, with dust lanes and dark patches. Coincident with many of
these features is extended 3.29--\um\ PAH emission, as shown in
Fig.~1(e). These features trace the PDRs and delineate the molecular
material. In addition, they wrap around the ionized material seen at 2\um. 

\subsection{The Galactic Centre}

Observations of the Galactic Centre obtained with SPIREX/Abu are far
less affected by the extinction than at optical and near--IR
wavelengths. The 4.05\,\um\, image in Fig.~1(f) clearly demonstrates
the advantages of this wavelength regime. Most of the 4.05\,\um\,
emission originates from heavily extincted sources, though there is
also some \bra\ emission, pinpointing regions of ionized gas. Three
sources, of roughly equal brightness, are prominent in the nucleus at
4\um. This is in contrast to the view at 2\um, which is dominated by
the source IRS~7.  These three sources are each separated by $\sim
7''$ and centred roughly on the presumed nucleus, Sgr\,A$^*$.  At the
spatial resolution achieved with SPIREX they are associated with the
sources IRS~1W, IRS~7/IRS~3 and IRS~13E/IRS~2L, respectively (as
imaged by Clenet et al.\ 2001, in the L--band).

\section{Conclusions}

Although SPIREX/Abu was a prototype facility, it nevertheless obtained
deep, high-resolution, wide-field images in the 3--5\um\ regime toward
many star forming complexes. These data are not only unique, but
when combined with complementary near--, mid--IR, and radio continuum
observations, reveal the inner environments of
star forming complexes (by tracing the PDRs), and identify the
youngest objects with circumstellar disks (through their high L--band
fluxes). The success of the SPIREX/Abu system lends strong support to
current plans to build larger IR telescopes on the Antarctic plateau,
where they would provide the most sensitive facilities for 3--5\,\um\,
observations on the Earth.

\section*{Acknowledgments}

The SPIREX project was a collaboration between the USA and Australia,
involving the National Optical Astronomy Observatories (NOAO), the
United States Naval Observatory (USNO), the Center for Astrophysical
Research in Antarctica (CARA), Boston University, Goddard Spaceflight
Center (GSFC), Ohio State University (OSU), Rochester Institute of
Technology (RIT), the University of Chicago (UC), the University of
New South Wales (UNSW), the Australian National University (ANU) and
the Universities Space Research Association (USRA).  We are indebted
to the dedicated efforts of our many colleagues here who together have
demonstrated that infrared astronomy can indeed be conducted from the
Antarctic plateau.


\end{document}